# Multimedia Steganographic Scheme using Multiresolution Analysis

Tirtha sankar Das
*Gurunanak Institute of Technology,
Kolkata-7000114*
tirthasankardas@yahoo.com

Ayan K. Sau, V. H. Mankar, Subir K. Sarkar
*Jadavpur University,
Kolkata-700032*
avankumarsau@yahoo.com. sksarkar@etce.idvu.ac.in

***Abstract:*** *Digital steganography or data hiding has emerged as a new area of research in connection to the communication in secured channel as well as intellectual property protection for multimedia signals. The redundancy in image representation can be exploited successfully to embed specified characteristic information with a good quality of imperceptibility. The hidden multimedia information will be communicated to the authentic user through secured channel as a part of the data. This article deals with a transform domain, block-based and signal non-adaptive/adaptive technique for inserting multimedia signals into an RGB image. The robustness of the proposed method has been tested compared to the other transform domain techniques. Proposed algorithm also shows improvement in visual and statistical invisibility of the hidden information.*

## 1. INTRODUCTION

The emergence of the World Wide Web as a popular means of communication has introduced new dilemmas regarding intellectual property [1]. From another point of view, different types of intended and non intentional signal impairments over the significant hidden information puts a limit on the data rate subject to a given embedding distortion [2]. One possible solution to this data communication and proprietary problem can be achieved through digital steganography/watermarking techniques [3,4]. Digital steganographic algorithms can be considered as digital communication scheme where auxiliary message (text, image, audio etc.) is embedded in digital multimedia signals and are available where ever the later signals move. In the present article, a block-based wavelet transform domain steganographic scheme is discussed where multimedia signals viz. text, image, audio etc. are redundantly embedded in homogeneity regions of RGB cover image [5,6]. The Discrete Wavelet transform (DWT) provides co-joint representation in the form of space frequency resolution. In the homogeneous blocks/regions, mean of the each block remains unchanged whereas variance is increased implying graceful degradation. The Daubechies-4 (db2 wavelet filter) provides a better trade off between the signal decomposition and time complexity for data hiding. The noisy grayscale version of the hidden multimedia information improves the data-embedding rate. The signal adaptive modulation function shows better robustness against common signal processing operations. The proposed algorithm shows better performance against copyright violation and at the same time, capacity of the cover image is highly enhanced [7]. The imperceptibility and robustness of this methodology has been tested against other transform domain techniques like DCT, Walsh/Hadamard, Fourier etc. The experimental result implies that the mutual information and probability of error are optimum and minimum respectively compared to the DCT and Walsh/Hadamard based data hiding techniques. This non-oblivious data-embedding scheme provides a better security of the hidden information.

## 2. WATERMARKING & DETECTION

An 8-bit gray scale image of dimension 256 X 256 is taken as background cover image. An 8-bit gray scale image of maximum $1/4^{th}$ of the cover image dimension, a 2 second single channel voice track (8KHz, 7kbps, 8Bit, Single Channel, wav format) and a text information written in notepad (Length maximum 256 characters) are taken as payload and to be transmitted in secret way.

### 2.1. Preprocessing of the Cover and signals

The gray scale cover is transformed to RGB colour space having three colour layers with same value. The image in colour space will look like gray scale image and the three layers will carry three different logos with no interference. Single channel 8-bit sound has values of signed integers ranging from –128 to 127, where as 8-bit gray scale image contains unsigned integers. So in order to embed the sound into the picture, it is converted into a two-dimensional unsigned gray scale image. Here, a bias value is used to change the negative values to positive ones because image pixel can't carry negative value. The text information is also converted into a binary image and then into the gray scale format. Now three types of information is converted into same type of gray scale image format.

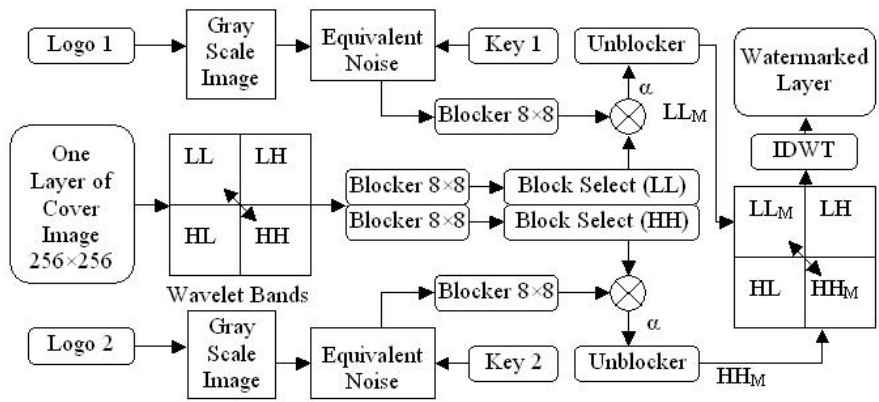

Fig 1. Block diagram of watermark insertion unit for a single layer of RGB image

### 2.2. Watermark insertion

Let I denotes the image block of dimension M×M i.e. image transformation coefficients of dimension M×M and L be the watermark information. The watermark W can be generated by performing the following operation:

$$W_M = \text{im2noise}(L, Key\_M\_<Key\ No.>) \quad (1)$$

The function *im2noise()* alters rows and columns randomly depending on the key used. The watermarked image $I_W$ can be obtained by embedding watermark information W into the image block I. The data embedding operation can be mathematically expressed as follows:

$$(I_W)_M = I_M + \alpha \bullet W_M \quad (2)$$

Where α is the modulation index and its proper choice will optimize the maximum amount of allowed distortion and minimum watermark energy needed for a reliable detection. α may or may not be a function of image coefficients.

### 2.3. Watermark detection

Being a non-oblivious data-embedding scheme the detection procedure is not a blind technique. This private key system requires the original cover image as well as the key set used during embedding technique. So,

$$W_M = (I_W)_M - I_M \quad (3)$$

$W_M$ is nothing but noise of inserted watermark. Logo L can be reconstructed by the reverse function of *im2noise*();

$$L = \text{noise2im}(W_M, Key\_M\_<Key\ No.>) \quad (4)$$

### 2.4. Discrete Wavelet Transform

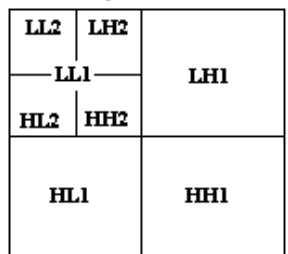

Fig 2. Wavelet sub bands

Discrete Wavelet transform provides co-joint representation in the form of space-frequency resolution, which in turn determines both local and global information of image signal [8]. The 2-band discrete wavelet system decomposes the image signal into LL, LH, HL and HH sub bands corresponding to different direction and resolution. In context of DWT domain watermarking, data embedding is recommended in wavelet coefficients of LL and HH sub bands because these two bands have minimum correlations.

**2.5. Image quality measure**

The *Structural Similarity Index Measurement* **(SSIM)** is an objective image quality assessment, which quantitatively measures and automatically predicts perceived image quality is defined as:

$$SSIM(x, y) = [l(x, y)]^{\alpha} \cdot [c(x, y)]^{\beta} \cdot [s(x, y)]^{\gamma} \qquad (5)$$

Where l(x,y), c(x,y) and s(x,y) are the luminance comparison, contrast comparison and structure comparison functions respectively and α>0, β>0 and γ>0 are parameters used to adjust the relative importance of the three components.

*Peak signal to Noise Ratio* **(PSNR)** is representative objective measure of data imperceptibility as this measure is popular and widely used in the literatures. The PSNR is expressed mathematically in the form as given below:

$$PSNR = (X_{max} \cdot Y_{max})I^2(X,Y)/\sum_{x,y}[I(X,Y) - I'(X,Y)] \qquad (6)$$

where I(x,y) represents a pixel value, with coordinates (x, y). The number of rows and columns in the pixel matrix is denoted by X and Y respectively.

Several researchers proposed various models to quantify *security*. In the present work, Cachin's security measure is used which is basically the relative entropy (Kulback Leibler distance) between the cover and watermarked (stego) images an mathematically given by:

$$D(P_X(x)P_R(x)) = \sum_{i=x} P_X(x) \log(P_X(x)/P_R(x)) \qquad (7)$$

Where random variables X and R represent the original and stego images respectively and $P_X(x)$ and $P_R(x)$ denote the probability mass function (PMFs) of X and R respectively. The lower is the value of $D(P_X(x)\| P_R(x))$, the better is the security and the value is always non-negative or zero. If $D(P_X(x)\| P_R(x)) =< \varepsilon$, the watermarking scheme is known as ε-secure, or the security value is said to be ε. If ε = 0, the system is known as perfectly secured system.

### 3. PROPOSED WATERMARKING SCHEME

Block diagram of the proposed watermarking scheme is given in the Fig.1. Two different watermarks (text-audio/audio-still image/still image-text etc.) are hidden in any one of the three colour layers of RGB cover image. The wavelet bands of cover image are partitioned into series of 8×8 blocks. This partitioning is also made for equivalent noise image of the watermarks i.e. logos are also partitioned into series of 8×8 blocks. Homogeneity regions of the

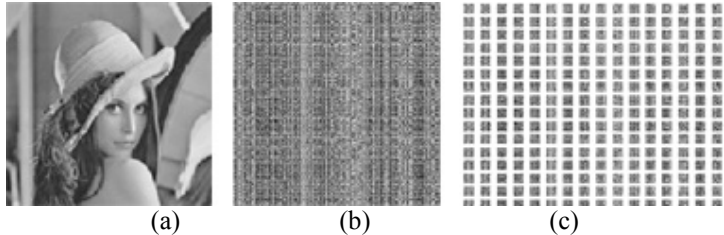

(a)　　　　　(b)　　　　　(c)
Fig.3. (a) Lena Image, (b) its noise equivalent, and (c) blocks for noise equivalent of Lena image.

wavelet coefficients of cover image are selected on the basis of variance values. Variance of each block is calculated for both LL & HH bands of each cover image layers. Watermarks are inserted in the homogeneous blocks arranged in ascending order (i.e. having small variance value) according to the equation (2) in LL & HH regions for Daubechies-4 (db2) wavelet filter coefficients. However other parameters can also be selected for adopting homogeneous regions. The remaining homogeneous blocks are replaced by an adaptive enhancement procedure. This effectively increases the variance of the blocks and ensures the robustness of the scheme. This procedure may lead to some blocking artifact effects in the watermarked image due to load in LL bands but no visible impact due to load in HH bands. Logos are distributed in the different bands of different layers as shown in table I. Each logo occurs twice one in LL band and other in HH band. This is done because LL and HH bands both are not sensitive to same attacks. Security of this scheme lies on the keys, original cover image, number of homogeneous blocks to carry the watermark and location of placement of logos.

| Layers | Logos | |
|---|---|---|
| | LL Band | HH Band |
| R | Text | Image |
| G | Image | Sound |
| B | Sound | Text |

Table I: Logo spreading

Detection procedure is the inverse operation of the insertion process. Being a non-oblivious watermarking scheme, original cover image is required in order to extract the logo. So to get back the watermarks, make wavelet transform of the cover image and also for three (R, G, B) layers of the watermarked image. Wavelet bands LL and HH are divided into 8x8 blocks. Calculate the variance of each block. Arrange the blocks according to the homogeneity criteria. Do the same for R, G and B three layers. Subtract data embedded homogeneous blocks from cover image blocks in order to get back the watermarked information. Reconstruct noise equivalent of logos by unblocking the series of extracted blocks. De-noise the noise equivalent watermarks using the same key being used during insertion process. Six watermarks will be reconstructed from R, G & B layers of the watermarked cover image, each twice. Optimizing from the obtained redundant watermarks, better detection scheme is achieved.

| Modulation index (m) | Colour Layer | PSNR value (dB) | SSIM value | Security ($\varepsilon$) value |
|---|---|---|---|---|
| 0.05 | R | 30.4541 | 0.9951 | 0.0658 |
| 0.05 | G | 30.2185 | 0.9969 | 0.0828 |
| 0.05 | B | 28.9071 | 0.9985 | 0.1245 |
| 0.10 | R | 26.47 | 0.9787 | 0.0513 |
| 0.10 | G | 28.41 | 0.9865 | 0.0498 |
| 0.10 | B | 29.28 | 0.9932 | 0.0645 |
| 0.15 | R | 22.7958 | 0.9490 | 0.0859 |
| 0.15 | G | 25.3432 | 0.9672 | 0.0552 |
| 0.15 | B | 26.9242 | 0.9833 | 0.0587 |

Table II: Results of PSNR, SSIM & Security value for Lena image using DWT decomposition

### 4. RESULTS AND DISCUSSION

We use PSNR and SSIM as representative objective measure of data imperceptibility where as relative entropy distance (Kullback Leibler distance) as measure of security ($\varepsilon$). Numerical results for these measures are shown in Table II. Structural similarity measure for the recovered logos tested and is noted in Table III. It is clear from the result that critical point appears near the modulation index 0.10 where tradeoff between cover image imperceptibility and recovered logo quality is achieved. However being an adaptive block based technique, this index may vary with different cover image. Result and blocking effect of the cover image may vary on choice of block size. Minimization in block size increases security value and time complexity but reduces blocking effect. This technique leads towards video frame watermarking for establishing a complete secured communication system establishment.

| Modulation index (m) | DWT Bands | SSIM for recovered Logos | | |
|---|---|---|---|---|
| | | Text | Image | Sound |
| m=0.05 | LL | 0.3038 | 0.3944 | 0.1198 |
| m=0.05 | HH | 0.7962 | 0.9924 | 0.9455 |
| m=0.10 | LL | 1.0000 | 0.9414 | 0.4494 |
| m=0.10 | HH | 1.0000 | 0.9990 | 0.9940 |
| m=0.15 | LL | 0.9470 | 0.9448 | 0.9130 |
| m=0.15 | HH | 1.0000 | 0.9986 | 0.9969 |

Table III: Structural similarity index measurement (SSIM) for recovered logos in different wavelet bands